# The influence of ultrafast temporal energy regulation on the morphology of Si surfaces through femtosecond double pulse laser irradiation

M. Barberoglou[1,2], G.D. Tsibidis[1,3]♣, D. Gray[1], E. Magoulakis[1,2], C. Fotakis[1,2], E. Stratakis[1], P. A. Loukakos[1],

[1] *Foundation for Research & Technology – Hellas, Institute of Electronic Structure and Laser P. O. Box 1527, Heraklion GR-71110, Greece.*
[2] *Department of Physics, University of Crete, Vassilika Vouton, GR-711 10, Heraklion, Greece.*
[3] *Materials Science and Technology Department, University of Crete, Heraklion 710 03, Greece.*

**ABSTRACT**

The effect of ultrashort laser-induced morphological changes upon irradiation of silicon with double pulse sequences is investigated under conditions that lead to mass removal. The temporal delay between twelve double and equal-energy pulses ($E_p$=0.24J/cm$^2$ each, with pulse duration $t_p$=430fs, 800nm laser wavelength) was varied between 0 and 14ps and a decrease of the damaged area, crater depth size and periodicity of the induced subwavelength ripples (by 3-4%) was observed with increasing pulse delay. The proposed underlying mechanism is based on the combination of carrier excitation and energy thermalization and capillary wave solidification and aims to provide an alternative explanation to the control of ripple periodicity by temporal pulse tailoring. This work demonstrates the potential of pulse shaping technology to improve nano/micro processing.

---

♣ Corresponding author: tsibidis@iesl.forth.gr

# 1. Introduction

Surface modification with ultrashort laser pulses has been studied extensively both in metal, semiconductors and dielectrics for many years [1-7] and several experimental and theoretical approaches have been proposed for the formation mechanisms of various structures [8-12]. Laser structuring on various materials is widely used for micromachining and it is of particular importance due to widespread applicability to a vast number of areas including integrated circuits, microfluidic chips and photovoltaics [13, 14]. A desirable effect in laser-mater processing applications is to influence in a controllable manner the morphology of the material surface by regulating the energy delivery from the laser into the various degrees of freedom of the system.

Temporally shaped femtosecond laser pulses have been employed to control thermal effects and improve micro/nano-scale material processing. Recent developments in optical devices have allowed any arbitrary shape of laser pulse to be generated [13]. Several studies include mainly experiments of laser ablation with double pulses as a first step towards understanding the effect of temporal pulse shaping in metals, semiconductors and dielectrics [15-22] . Theoretical simulations relating to the main stages of the double pulse ablation process in metals were performed at different time delay windows and conditions [23]. In most approaches, evaporation and ablation effects are assumed to induce a remarkable surface modification. Depending on the experimental parameters (pulse duration, fluence, pulse number, etc), a variety of morphological changes are expected, ranging from simple craters to ripples and micro/nano structures [24, 25]. With respect to the formation of ripples, various mechanisms have been proposed to account for the formation of periodic structures: interference of the incident wave with an induced scattered wave [26, 27], or with a surface plasmon wave (SPW) [25, 28], or due to self-organisation mechanisms [29]. Recently, a theoretical framework was proposed that elaborates on both optical (the interference of the incident and plasmon waves) and hydrodynamical (capillarity-driven ripple formation) effects to account for surface modification and ripple periodicity after irradiation with multiple pulses [30].

Although a previously unexplored area on surface modification (i.e. crater and ripple formation) due to double pulse laser irradiation on semiconductors under conditions that induce mass removal has only recently started to be investigated [16, 22, 31], the particular role of the dynamics of a superheated liquid material in the morphological changes still needs to be addressed. In a recent study on irradiation with single pulses, hydrodynamics was proposed as an alternative to previous scenarios [25, 32] and dependence of the ripple periodicity with increasing number of pulses demonstrated a very good agreement with experimental observations [30]. In the case of single pulses, the associated thermal collateral effects induce larger spot volume changes (i.e. crater depth and size) which can have an adverse effect. By contrast, the possibility of controlling the ripple periodicity through manipulation of the temporal energy delivery and reducing the residual damage is a very important issue to explore as it would offer unique capabilities for improving micro/nano-processing. To examine the contribution of the hydrodynamical factor in the rippled structures with increasing number of temporally separated pulses, a thorough investigation of the physical fundamentals of the combined heat transfer and hydrodynamics modules would be required. Furthermore, elucidation of the underlying mechanisms would allow to explore the interplay of the optical (i.e. wave interference) and hydrodynamical roles in determining the ripple periodicity based on recent experimental findings which showed that in the case of irradiation with double pulses, ripple periodicity in silicon is rather independent of the pulse delay [22].

Hence, in the present work, we aim to present a complementary experimental and theoretical investigation of irradiation of silicon with double pulses as a basic form of temporally shaped pulses in ablation conditions. The laser energy of double pulses with various time separations is delivered in a time window pertinent to the electron–lattice interaction. To offer an insight into the modification process and elucidate the underlying mechanism of the morphology alteration, a hybrid theoretical model is introduced that takes into account ultrafast heat transfer from carriers to the lattice and hydrodynamics during the solid-liquid-solid phase transition. The morphological changes (i.e. damaged area, crater depth, ripple periodicity) are examined with respect to the temporal delay between the individual pulses that comprise the double pulse sequence and a comparison between theoretical results and experimentally obtained observations is conducted to validate the proposed mechanism.

## 2. Experimental

Experiments were performed with a femtosecond Ti:Sapphire laser system operating at a wavelength of 800 nm and repetition rate of 1 kHz. The pulse duration was set to 430 fs and measured by means of an auto correlation technique. A 4/F pulse shaper configuration using a Spatial Light Modulator (SLM) was used in order to filter the Fourier spectrum of the laser pulses and create double pulse sequences of equal fluence (0.24 J/cm$^2$ each) with pulse separations, $t_d$, varying from 0 to 14ps. A pockels cell controlled the repetition rate and the number of the double pulse sequences, $NP$, that irradiated the silicon surface. Single crystal n-type (phosphorous-doped) Si (1 0 0) polished wafers (Siltronix) with a resistivity of 2-8 Ohm·cm were used. The irradiation took place in a vacuum chamber evacuated down to a residual pressure of ~10$^{-2}$ mbar by means of a rotary pump (Alcatel).The beam was subsequently focused with a quartz lens (*f*=5cm) on the sample, mounted to a sample holder inside the vacuum processing chamber giving a spot diameter of 15μm. The laser beam entered the chamber through a quartz entrance window, while the irradiation process could be monitored through a Plexiglas window, which was laterally mounted on the vacuum chamber. The processing chamber was placed on a computer driven high precision X-Y translation stage (Standa) with spatial resolution of 1 μm. The maximum energy delivered onto the silicon target was 1.2 μJ per pulse corresponding to a laser fluence of 0.48 J/cm$^2$. Various numbers of shots were used ranging from 10 – 1000 laser shots. Field emission scanning electron microscopy (FESEM) was used for imaging analysis and spot diameter characterization. Furthermore, atomic force microscopy (AFM) was used to evaluate the crater depth profiles of several spots at different time delays. In this experimental scheme, by introducing spatial filters into the Fourier spectrum of a laser pulse the temporal shape of the laser pulse is manipulated.

## 3. Theoretical framework

Ultrashort-pulsed lasers first excite the charge carriers (electron-hole pairs) in semiconductors while their energy is subsequently transferred to the lattice. The relaxation time approximation to Boltzmann's transport equation is employed to determine the number density, carrier energy and lattice energy [33]. The evolution of the number density $N$, carrier temperature $T_c$ and lattice temperature $T_l$ are derived using the corresponding equations. Based on this picture the following set of equations determine the temperature and particle dynamics [33-35]

$$C_c \frac{\partial T_c}{\partial t} = \vec{\nabla} \cdot \left( (k_e + k_h) \vec{\nabla} T_c \right) - \frac{C_c}{\tau_e} (T_c - T_l) + S(\vec{r},t)$$

$$C_l \frac{\partial T_l}{\partial t} = \vec{\nabla} \cdot \left( K_l \vec{\nabla} T_l \right) + \frac{C_c}{\tau_e} (T_c - T_l)$$

$$\frac{\partial N}{\partial t} = \frac{\alpha}{h\nu} \Omega I(\vec{r},t) + \frac{\beta}{2h\nu} \Omega^2 I^2(\vec{r},t) - \gamma N^3 + \theta N - \vec{\nabla} \cdot \vec{J} \qquad (1)$$

$$\Omega = \frac{1 - R(T_l)}{\cos \phi}$$

where $C_c$ and $C_l$ are the heat capacity of electron-hole pairs and lattice, respectively, $\nu$ is the frequency of the laser beam corresponding to a wavelength equal to $\lambda$=800nm, $\varphi$ is the angle between the vertical axis and the beam direction, $k_e$ and $k_h$ are the thermal conductivity of the electrons and holes, respectively, $K_l$ is the thermal conductivity of the lattice, $h$ is the Plank's constant, $\gamma$ is the Auger recombination coefficient, $\theta = 3.6 \times 10^{10} \, e^{-1.5 E_g / k_B T_c}$ sec$^{-1}$ is the impact ionisation coefficient, while $\alpha$ =0.112 $e^{T_l/430}$ μm$^{-1}$ and $\beta$=9×10$^{-5}$ sec μm/J are the one-photon and two-photon absorption coefficients, respectively, $R(T_l)$ is the reflectivity of the laser beam on the silicon surface (=0.329+5×10$^{-5}$ $T_l$), $\tau_e$ is the energy relaxation time (=1ps), $\vec{J}$ is the carrier current vector and $S(\vec{r},t)$ is provided by the following expression

$$S(\vec{r},t) = (\alpha + \Theta N) \Omega I(\vec{r},t) + \beta \Omega^2 I^2(\vec{r},t) - \frac{\partial N}{\partial t}(E_g + 3k_B T_c) - N \frac{\partial E_g}{\partial T_l} \frac{\partial T_l}{\partial t} - \vec{\nabla} \cdot \left( (E_g + 4k_B T_c) \vec{J} \right) \qquad (2)$$

where $\Theta$ stands for the free-carrier absorption cross section (=2.9×10$^{-10}$ $T_l$ / $T_0$ μm$^2$), $k_B$ is the Boltzmann's constant and $E_g$ the band-gap energy (a full list of the values of the parameters used in this model are in Ref. [30]). For the sake of simplicity, it is assumed that the incident beam is normal to a flat irradiated surface (hence $\varphi$=0 for the first pulse, while for subsequent pulses $\varphi \neq 0$). The laser intensity in Eqs.(1) is obtained by considering the propagation loss due to one-, two- photon and free carrier absorption

$$\frac{\partial I(\vec{r},t)}{\partial z} = -(\alpha + \Theta N) I(\vec{r},t) - \beta I^2(\vec{r},t) \qquad (3)$$

assuming that the laser beam is Gaussian both spatially and temporally, and the transmitted laser intensity at the incident surface is expressed in the following form

$$I(r, z=0, t) = \frac{2\sqrt{\ln 2}}{\sqrt{\pi} \tau_p} \frac{E_p}{2} e^{-\left(\frac{2r^2}{R_0^2}\right)} \left( e^{-4\ln 2 \left(\frac{t-t_0}{\tau_p}\right)^2} + e^{-4\ln 2 \left(\frac{t-t_0-t_{delay}}{\tau_p}\right)^2} \right) \qquad (4)$$

where $t_{delay}$ corresponds to the temporal separation between the two pulses. It is assumed that the total laser beam fluence is equally shared by the two constituent pulses. $E_p$ stands for the fluence of the laser beam and $\tau_p$ is the pulse duration (i.e. full width at half maximum), $R_0$ is the irradiation spot-radius (distance from the centre at which the intensity drops to $1/e^2$ of the maximum intensity); $t_0$ is chosen to be equal to $2\tau_p$. The choice of the value of $t_0$ is based on the requirement that at $t=0$, the intensity of the incident beam is practically zero, at $t=t_0$ it reaches the maximum power while laser irradiation practically vanishes at $t=2t_0$.

The experimental conditions that are investigated are expected to induce material removal. To introduce material loss in the theoretical framework, the assumption is made that all lattice points beneath the surface that undergo explosive boiling are removed from the material along with the associated thermal energy. A consistent analysis though, should involve elimination of the subset of the electron system that during the electron-lattice heat exchange leads to temperatures greater than ~$0.90T_{cr}$ [8, 30, 36, 37] (for silicon, $T_{cr}$ =5159$^0$K). To describe the heat transfer in the superheated material that remains in the system, a revised two-temperature model that describes heat transfer from electrons to lattice has to be employed [38] taking into account that the molten material exhibits metal behaviour. Hence, for temperatures above $T_m$ (~1687 $^0$K), Eqs.1 needs to be replaced by the following two equations that describe electron-lattice heat transfer

$$C_e \frac{\partial T_e}{\partial t} = \vec{\nabla} \cdot \left( K_e \vec{\nabla} T_e \right) - \frac{C_e}{\tau_E}(T_c - T_L)$$
$$C_L \frac{\partial T_L}{\partial t} = \frac{C_e}{\tau_E}(T_c - T_L) \qquad (5)$$

where $C_e$ and $C_L$ are the heat capacity of electrons and lattice (liquid phase), $K_e$ is the thermal conductivity of the electrons, while $\tau_E$ is the energy relaxation time for the liquid phase. Furthermore, to describe the phase change in the interface between liquid and solid phase the second equation in Eq.1 has to be modified properly to include the phase transition

$$\left( C_l \pm L_m \delta(T_l - T_m) \right) \frac{\partial T_l}{\partial t} = \vec{\nabla} \cdot \left( K_l \vec{\nabla} T_l \right) + \frac{C_c}{\tau_e}(T_c - T_l) \qquad (6)$$

where $L_m$ is the latent heat of fusion and

$$\delta(T_l - T_m) = \frac{1}{\sqrt{2\pi}\Delta} e^{-\left[\frac{(T_l-T_m)^2}{2\Delta^2}\right]} \qquad (7)$$

where $\Delta$ is in the range of 10-100$^0$K depending on the temperature gradient.

The superheated material which is in the liquid phase behaves as an incompressible Newtonian fluid and flow and heat transfer in the molten material are defined by the following equations:

(i). for the mass conservation (incompressible fluid):

$$\vec{\nabla}\cdot\vec{u}=0 \tag{8}$$

(ii). for the energy conservation

$$C_L\left(\frac{\partial T_L}{\partial t}+\vec{\nabla}\cdot(\vec{u}T_L)\right)=\vec{\nabla}\cdot(K_L\vec{\nabla}T_L) \tag{9}$$

where $K_L$ is the thermal conductivity of the lattice. The presence of a liquid phase and liquid movement requires a modification of the second of Eq.1 to incorporate heat convection. Furthermore, an additional term is presented in the equation to describe a smooth transition from the liquid-to-solid phase (i.e. it will help in the investigation of the resolidification process)

$$C_L\left[\frac{\partial T_L}{\partial t}+\vec{\nabla}\cdot(\vec{u}T_L)\right]-L_m\delta(T_L-T_m)\frac{\partial T_L}{\partial t}=\vec{\nabla}\cdot(K_L\vec{\nabla}T_L) \tag{10}$$

(iii). for the momentum conservation:

$$\rho_L\left(\frac{\partial \vec{u}}{\partial t}+\vec{u}\cdot\vec{\nabla}\vec{u}\right)=\vec{\nabla}\cdot\left(-P\mathbf{1}+\mu(\vec{\nabla}\vec{u})+\mu(\vec{\nabla}\vec{u})^T\right) \tag{11}$$

where $\vec{u}$ is the velocity of the fluid, $\mu$ is the liquid viscosity, $P$ pressure. $C_L$ and $K_L$ stand for the heat capacity and thermal conductivity of the liquid phase, respectively. It is evident that the transition between a purely solid to a completely liquid phase requires the presence of an intermediate zone that contains material in both phases. In that case, Eq.10 should be modified accordingly to account for a liquid-solid two phase region (i.e. mushy zone) where the total velocity in a position should be expressed as a combination of the fraction of the mixtures in the two phases [39]. Nevertheless, to avoid complexity of the solution of the problem and given the small width of the two phase region with respect to the size of the affected zone a different approach will be pursued where a mushy zone is neglected and transition from to solid-to-liquid is indicated by a smoothened step function of the thermophysical quantities (see [30] for a detailed description).

Vapour ejected creates recoil pressure on the liquid free surface which pushes the melt away in the radial direction. The recoil pressure and the surface temperature are usually related according to the equation [40, 41]

$$P_r=0.54P_0\exp\left(L_v\frac{T_L^S-T_b}{RT_L^S T_b}\right) \tag{12}$$

where $P_0$ is the atmospheric pressure (i.e. equal to $10^5$ Pa), $L_v$ is the latent heat of evaporation of the liquid, $R$ is the universal gas constant, and $T_L^S$ corresponds to the surface temperature. Given the radial dependence of the laser beam, temperature decreases as the distance from the centre of the beam increases; at the same time, the surface tension in pure molten silicon decreases with growing melt temperature (i.e

*dσ/dT<0*), [40] which causes an additional depression of the surface of the liquid closer to the maximum value of the beam while it rises elsewhere. Hence, spatial surface tension variation induces stresses on the free surface and therefore a capillary fluid convection is produced whereas a small protrusion is formed near the edge of the spot. Moreover, a precise estimate of the molten material behaviour requires a contribution from the surface tension related pressure, $P_\sigma$, which is influenced by the surface curvature and is expressed as $P_\sigma = K\sigma$, where $K$ is the free surface curvature. The role of the pressure related to surface tension is to drive the displaced molten material towards the centre of the melt and restore the morphology to the original flat surface. Thus, pressure equilibrium on the material surface implies that the pressure in Eq.10 should outweigh the accumulative effect of $P_r + P_\sigma$. Due to an anticipated variation in the amount of ablated region upon irradiation in different ambient conditions, the associated recoil pressure is expected to be dependent on the environment. Hence, a systematic study of recoil pressure dependence is required to reveal the contribution to morphological modification in various ambient conditions, however, an investigation is beyond the scope of the present work.

As the material undergoes a solid-to-liquid-to-solid phase transition, it is important to explore the dynamics of the distribution of the depth of the molten material and the subsequent surface profile change when solidification terminates. The generated ripple height is calculated from the Saint-Venant's shallow water equation [42]

$$\frac{\partial H(\vec{r},t)}{\partial t} + \vec{\nabla} \cdot \left( H(\vec{r},t)\vec{u} \right) = 0 \qquad (13)$$

where $H(\vec{r},t)$ stands for the melt thickness. Hence, a spatio-temporal distribution of the melt thickness is attainable through the simultaneous solution of Eqs.(1-13).

Due to the axial symmetry of the laser beam, a two-dimensional approach is pursued to solve the aforementioned set of equations and describe the process that leads to a surface modification upon laser irradiation with a single double pulse. Further exposure to multiple pulses requires consideration of the interference of the incident wave with a SPW which results in a spatial and periodic modulation of the energy that is deposited onto the materials spatially modified surface. The involvement of a plasmon wave related mechanism in the generation of ripples was employed as the metallic behaviour of silicon at high temperatures allows excitation of SPW. The plasmon wavelength, $\lambda_s$, is related to the wavelength of the incident beam through the relations and the dielectric permittivity $\varepsilon'$ [25]

$$\lambda_s = \lambda \left( \frac{\varepsilon' + \varepsilon_d}{\varepsilon' \varepsilon_d} \right)^{1/2}$$

$$\varepsilon' = \mathrm{Re}\left( 1 + (\varepsilon_g - 1)\left(1 - \frac{N}{n_0}\right) - \frac{N}{N_{cr}} \frac{1}{\left(1 + i\frac{1}{\omega\tau_e}\right)} \right) \qquad (14)$$

where $\varepsilon_d$ ($\varepsilon_d = 1$) is the dielectric constant of air, $\varepsilon_g$ stands for the dielectric constant of unexcited material ($\varepsilon_g=13.46+i0.048$), $\omega$ is the frequency of the incident beam, $n_o$ is the valence band density ($n_o=5\times10^{22}$cm$^{-3}$), $N_{cr}=m_{eff}\varepsilon_0\omega^2/e^2$ where $m_{eff}$ is the effective electron mass (i.e. $1.08m_e$ in Si). As the number of pulses increases, the ripple peaks become more pronounced and the increased profile curvature hinders the overall heat absorption (due to the change of the local angle of incidence [32]) which, in turn, leads to a decrease of the carrier temperature peak and density of excited carriers. Hence, Eq.14 infers that the wavelength of the surface plasmon wave and subsequently the ripple periodicity drops with decreasing number of carriers for an increasing number of pulses. An alternative explanation has been suggested in which the increased deepening of the grating like surface relief leads to a lower resonant wavelength of the SPW [25]. Periodic structures (i.e. ripples) will be created upon melting-solidification with an orientation perpendicular to the polarization of the electric field. Furthermore, the interference of the initial beam with the SPW will destroy the axial symmetry of the system and thereby surface morphology will not exhibit a cylindrical symmetry. As a result, for $NP \geq 2$, a three dimensional solution of Eqs.(1-13) is employed while simulation takes into account the new profile that is created after the surface modification at the end of solidification procedure (see Ref [30] for a detailed description).

## 4. Results and discussion

A primitive form of pulse shaping was performed by splitting the initial pulse into two equal components and varying the temporal pulse separation in some few picoseconds time range thus attempting to manipulate the ultrafast electron interaction and cooling processes that take place in this temporal range for many solids in general and in silicon in particular. Crystalline silicon was irradiated using multiple laser double pulses (up to 1000 pulses), and different $t_d$ ranging from 0 to 14ps. Fig.1 shows the spot size variation as a function of number of pulses (*NP*) and $t_d$. The SEM images indicate a spot area decrease with increasing $t_d$ for a fixed pulse number. Quantitative analysis of the spot area details provided by SEM images showed a pronounced size decrease with increasing $t_d$, as illustrated in Fig.2. By contrast, the experimental observations (Fig.2) also demonstrate a profound increase of the spot area size with *NP*, at constant $t_d$, which also has been observed for single pulses [30] (in order to emphasise that the spot area decrease only results for $t_d$ ranging from 0 to 2ps are illustrated in Fig.2).

To obtain a detailed picture of the surface morphology by analyzing results of ripple frequency and spot depth, AFM images were taken for *NP*=12 and the spot area was observed for each $t_d$. It should be noted, though, that the intensity-based profile (SEM images) provides accurate information only for the dimensions in the sample plane, while the vertical direction it gives relative rather than the absolute dimensional values. A depth-related analysis approach through processing of the AFM image and plotting the variation of the depth across selected cross sections along the direction perpendicular to the orientation of the electric field of the laser beam (*black* line in Fig.3a) shows that the peaks of the ripples grow above the flat surface (Fig.3b). Fig.3a displays a representative image of the spot formed after irradiation with 12 single pulses (i.e. no double pulse sequences). The ripple periodicity was calculated by applying a Fast Fourier Transform (FFT) yielding an average value equal to 747±12nm (Fig.3b). Fig. 3c presents the corresponding dependence of the ripples periodicity on $t_d$, showing a distinguishable decrease of the ripples wavelength with increasing pulse separation.

For higher *NP*, ripples in the center of the spots are replaced by bigger structures with size of a few micrometers, while a depression is also observed at the spot center. Furthermore, as shown in Fig. 4 (*a* to *c*), comparing the SEM images of the craters formed for single and double pulse irradiation, the crater depth decreases upon increasing $t_d$ . We further analyzed, via AFM imaging, this crater suppression, which is directly related to the double pulse irradiation. . Figure 4d,e illustrates a contour analysis of the depth profile obtained from the AFM images for spots irradiated with 12 pulses for zero and 0.6ps pulse delays. It is evident, that the crater size decreases which confirms the initial observation by SEM. A profile analysis (Fig.4d,e, inset) across the center of the beam provides more details about the morphology of the crater. In particular, for a big overlap between the two component pulses of the train (i.e. $t_d$=0), ripples are formed below the level of the flat surface; by contrast, for longer delays (i.e. $t_d$=600 fs), the ablated region is smaller and the peaks of the ripples emerge above the flat surface. A comparison between the ripple areas of the same spot (Fig.3a and Fig.4d) demonstrate that ablation is evident close to the centre of the maximally heat affected region as the energy deposition is extremely high. However, at the periphery of the spot where the fluence is smaller, the ripple profile follows a common interference pattern with the observed peaks for the case of Fig.3a where the ripples extend above the flat surface.

In order to evaluate the morphological changes as a consequence of the ultrafast primary mechanisms that occur in silicon following laser excitation, we will discuss them along with our calculations of the maximum surface carrier and lattice temperatures $T_{c,max}$ and $T_{l,max}$ respectively, that are shown in Fig.5. Fig.5 illustrates the dependence of the maximum surface carrier and lattice temperatures as functions of the combined effect of both pulses in the double pulse sequence for *NP*=12. The biggest value of the $T_{c,max}$ is reached for the maximum temporal overlap of the double pulses and it decreases monotonically for increasing interpulse delay as shown with the dashed blue line in Fig 5. The form of the dependence of $T_{l,max}$ is more complicated and should be explained in terms of the interplay between two competing mechanisms: carrier transport which transfers heat away from the laser excited region and the carrier-lattice coupling which induces heat localization.

To help us evaluate the impact of the pulse separation on the lattice temperature, we consider two ranges of pulse separation: (a) $t_{delay}$<$5\tau_p$ (range I) (b) $t_{delay}$≥$5\tau_p$ (range II). More specifically, for interpulse delays in range I, $T_{c,max}$ is high, thus inducing an efficient energy transport away form the surface region into the bulk (due to the high carrier conductivity). Therefore, less energy remains at the surface region available to be transferred to the surface lattice region by carrier-lattice collisions. On the other hand, as the interpulse delay is further increased and $T_{c,max}$ decreases drastically (Regime II), the channel of energy transport due to the carrier diffusion weakens. Thus, more energy in the surface region is available to be transferred toward the lattice before it is dissipated into the bulk through transport mechanisms. Hence $T_{l,max}$ reaches its maximum value for longer interpulse delays than $T_{c,max}$. A further increase of the pulse separation would be followed by a slow decrease of the lattice temperature as the surface temperature due to the first pulse is falling off when the second pulse reaches the surface. This is also indicated by the absence of overlapping of the two carrier temperature profiles (Fig.6). The optimum pulse separation that corresponds to the maximum $T_{l,max}$ is almost $5\tau_p$ for the conditions assumed in the simulations. An initial increase of the surface temperature of the material has been previously observed in metals in double pulse experiments [10, 43], which we also expect for silicon as in very high temperatures the semiconductor undergoes a transition to a metallic state.

Based on the above discussion, we can interpret the dependence of the morphological characteristics, namely the crater area and the crater depth shown in Fig. 7a,b in terms of $T_{c,max}$ and $T_{l,max}$. For increasing interpulse delay, the crater depth and spot area of the material that has undergone mass removal and phase transition are smaller; as seen in both figures the decrease of the morphological characteristics is monotonic and starts for zero interpulse delay, therefore exhibiting a similar behavior to the $T_{c,max}$ instead of $T_{l,max}$. Therefore we conclude that the particular ablation process for our experimental conditions shows a stronger dependence on the carrier dynamics rather than the lattice temperature dynamics. The predominant role of the carrier dynamics and diffusion in morphological changes has also been emphasized in previous works [3, 44] while a recent theoretical approach explained the decrease of the crater depth by the suppressing character of the interaction of the shock wave produced by the second pulse and the rarefaction wave of the first pulse [23].

Looking at the influence of the pulse number on the ablation process we note that an increase of the number of pulses for the same time delay results in larger spot sizes (Fig.1) as the total local energy density is higher and diffusion of energetic carriers penetrate the material deeper and increases the size of the heat affected areas (Fig.2). The investigation of the physical mechanisms that induce surface modification, for $NP>1$, requires the consideration of the combination of optical effects (interference of SPW with the incident wave), thermalisation and hydrodynamics. To incorporate the contribution of hydrodynamics, we need to emphasise that due to the total periodic deposition of the laser density (resulting from the interference of the incident wave and the SPW), the recoil pressure produced will also be a periodic function and spatially modulated [30]. Nevertheless, the values of local maxima decrease towards the edge of the affected region which indicates that at positions where the recoil pressure function has local maxima, upward flow will be stronger on the side that is closer to the centre. Fig.7c shows the spatial dependence of the dynamics of the molten material results for $NP$=12 at $t$=1ns for a single pulse where the arrows indicate the flow direction. The electric field of the laser beam is assumed to be polarised along the $x$-axis while the position $(x,y,z)$=(0,0,0) corresponds to the point where the intensity of the laser beam is highest.

Fig.7a,b illustrate a comparison of the predicted theoretical results and data from experimental observations for the spot area and crater depth as a function of the pulse delay for $NP$=12. For small pulse delays, the spot area and crater depth decrease rapidly where there is a slower decrease for bigger pulse delay. Thus, it is evident that similar morphological changes (i.e. for spot area and crater depth) occur regardless of the number of pulses. Furthermore, it is shown that the steep decrease of the spot area and the crater depth follows the trend of the maximum carrier temperature (Fig.5) which demonstrates the predominant role of the carrier dynamics in the resulting morphological changes. A similar decreasing pattern was observed for different conditions in other studies where the initial fast decrease is attributed to the role of the multi-photon absorption in the carrier generation which is more effective for energy deposition with shorter pulse delays [16]. While Fig.7a shows an adequate agreement between the theoretical and experimental results for the spot area dependence on pulse separation, it appears that the theoretical model provides an overestimation for the maximum spot depth (Fig.7b) for longer delays which can be explained by a less precise experimental estimation (i.e. measurement) of small depth sizes. Fig.7d illustrates the spatial distribution of the lattice temperature field at $t$=1ns for 11 double pulses (with no time separation between the two pulses of each train) which indicates the spatial periodicity of the lattice temperature as a result of the interference of the incident and the SP waves.

Fig.8a shows the comparison of the periodicity of the subwavelength ripples as a function of the pulse separation for $NP$=12. The decrease of the periodicity with increasing pulse delay is due to the decrease of the electron temperature and carrier density which according to Eq.14 leads to smaller values of the associated SPW. It is evident that despite some discrepancy between the theoretical and experimental results, especially for longer pulse separation, there is a satisfactory prediction of the ripple values and periodicity change. The conspicuous deviation of the wavelengths with (*dashed* line) and without (*solid* line) the incorporation of the recoil pressure contribution indicates the significance of the role of the recoil pressure and Marangoni flow; Due to the local increase of the deposited energy at smaller distances from the centre, recoil pressure will push the material deeper and away from centre, however, pressure due to surface tension will increase gradually balancing the contribution of recoil pressure at longer times ([30]). Thus, two competing forces (i.e. recoil and surface tension) will squeeze further the originally produced profile (with spatial periodicity equal to the surface plasmon wavelength) and therefore the ripple wavelength is decreasing. Moreover, the theoretical results show that hydrodynamics and phase change process (i.e. solidification mechanism) considerations do not lead to suppression of the ripples.Fig.8a also shows that ripple wavelength decrease with increasing number of pulses through the consideration of the recoil pressure (besides the expected decrease of SPW) yields a better agreement with our experimental observations. In comparison with previous experimental results under similar conditions that showed an almost independent on the double-pulse delay [22], the proposed theoretical model predicted a deviation of the produced subwavelength ripples by 3-4% over the entire range of time delays. Furthermore, the experimental observations used to validate the proposed theoretical framework showed a consistency within the uncertainties of both measurements (i.e. current and previous study).

It is important to emphasise that one could argue that any combination of fluence and number of double pulses could be employed to modulate the morphological characteristics upon irradiation with double pulses. Nevertheless, this assumption is not accurate from a theoretical point of view and it is upon experimental confirmation. More specifically, as the number of pulses increases, the crater depth increases due to a higher constrained energy deposition while a subsequent hydrodynamical movement leads to more pronounced hills and ripples. In the present work, the combination of $E_p$=0.24J/cm$^2$ (for each pulse) and $NP$≤22 led to a simultaneous increase of crater depth and ripple height. However, for higher NP, the magnitude of the steepness of the hill is large enough to lead to big lattice temperature gradient while part of the hill fails to reach the melting temperature. As a result, hydrodynamical movement will be leading to an increasingly fluctuating profile. To illustrate the behaviour, a simulation was conducted for irradiation with a double pulse ($E_p$=0.24J/cm$^2$ for each pulse, $NP$=23, $t_d$=100fs) (Fig.8b). Similar effect is observed also for a combination of higher fluence and lower number of pulses. More specifically, for $E_p$≥1.23J/cm$^2$ (for each pulse) and $NP$=12 ($t_d$=100fs), a non smooth ripple profile is derived. Further investigation, though, of the aforementioned conditions and associated effects is required both on the computational level (i.e. that lead to fluctuating computational errors) or on the need to incorporate an improved physical mechanism.

In conclusion, the above investigation demonstrates that ripple periodicity is controllable by tailoring the temporal shape of individual pulses while the proposed combined mechanism of carrier excitation and dynamics of the superheated material (solid to liquid to solid transition) provides an efficient interpretation of the surface modification. Although a decrease of ripple periodicity is feasible by increasing the number of pulses and using single pulses, a repetitive exposure of the material to laser heating results in an enhanced residual damage as every pulse produces a new deeper profile [30]. Furthermore, increase of the energy deposition by simply increasing the single's pulse fluence limits the window range to avoid unwanted effects (for example, excessive ablation). Hence, in some applications it may be preferable to use the double pulse technology to modulate periodicity and morphological changes by keeping the total energy constant than vary the deposited energy. By contrast, the above analysis demonstrates the uniqueness of the pulse delay methodology to control ripple periodicity with a simultaneous heat localisation and decreased material removal (i.e. smaller volume removed) which can be important in a number of applications in modern technology and biomedicine. Furthermore, the theoretical framework manages to describe adequately the correlation of the size and depth of the ablated region with the pulse separation and the presence of an optimum pulse separation value (corresponding to a maximum surface lattice temperature) which can be the basis for the production of well-defined micro/nanostructures with high efficiencies.

## 5. Conclusions

The experimental and theoretical results presented in this work: (a) provide a satisfactory insight into the process that dictates the morphological changes and formation of ripples upon irradiation of silicon surfaces with ultrashort double pulses, (b). shed light on the underlying mechanisms and associate the related morphological changes to a combination of carrier excitation and hydrodynamic factors. The rigorous and systematic analysis of the induced morphological changes has led to further elucidation of the connection between the advantages of the pulse shaping technology (such as ripple periodicity control) and characteristics of the produced spot; hence, the approach could further allow control of the opto-electronic properties of materials by modifying the temporal characteristics of the laser processing beam.

**Acknowledgements**

This work was supported by the Integrated Initiative of European Laser Research Infrastructures *LASERLAB-II* (Grant Agreement No. 228334). G.D.T and E.Stratakis acknowledge financial support from the *'3DNeuroscaffolds'* research project.

Figure 1: SEM image of silicon samples irradiated with various number of double pulses and delays ($E_p$=0.24J/cm$^2$ per pulse, $t_p$=430fs, 800nm laser wavelength).

Figure 2: Experimental data for spot area dependence on pulse delay as a function of the number of laser pulses per spot ($E_p$=0.24J/cm$^2$ per pulse, $t_p$=430fs, 800nm laser wavelength).

Figure 3: (a) AFM image of spot irradiated with 12 single pulses, (b) Intensity profile of surface roughness across a line passing through the ripples-dashed line represent intensity profile on flat surface. (c) Ripple periodicity as a function of pulse separation for $NP$=12, ($E_p$=0.24J/cm$^2$ per pulse, $t_p$=430fs, 800nm laser wavelength).

Figure 4: (a) to (c): Suppressions of the crater formed upon irradiation of the silicon target with 1000 pulses for pulses delays 0ps, 0.5ps, and 2 ps, respectively. (d). Contour analysis of AFM images from the spots irradiated with 12 pulses for zero pulse delay, and (e) and 0.6ps. Insets illustrate a mean value of the crater depth, ($E_p$=0.24J/cm$^2$ per pulse, $t_p$=430fs, 800nm laser wavelength).

Figure 5: Irradiation with $NP$=12: Maximum lattice and electron temperature dependence on pulse separation ($E_p$=0.24J/cm$^2$ per pulse, $t_p$=430fs, 800nm laser wavelength).

Figure 6: Electron temperature for pulse separation equal to $5t_p$ (inset shows the pulse disengagement for separation equal to $5t_p$). ($E_p$=0.24J/cm$^2$ per pulse, $t_p$=430fs, 800nm laser wavelength).

Figure 7: Morphological changes: (a) spot area *vs.* pulse delay ($NP$=12), (b) maximum crater depth size *vs.* pulse delay ($NP$=12), (c) spatial dependence of flow pattern (indicated by the arrows) $t$=1ns (single pulse, $NP$=12), and (d) Spatial distribution of lattice temperature, at $t$=1ns for $NP$=11 (single pulse). ($E_p$=0.24J/cm$^2$ per pulse, $t_p$=430fs, 800nm laser wavelength).

Figure 8: For $E_p$=0.24J/cm$^2$ per pulse, $t_p$=430fs, 800nm laser wavelength: (a) ripple periodicity as a function of pulse delay for $NP$=12 (Theoretical results are presented in absence and with the presence of recoil pressure (*RP*) contributions). (b) Ripple profile for $NP$=23 ($t_d$=100fs).

**Figure 1**

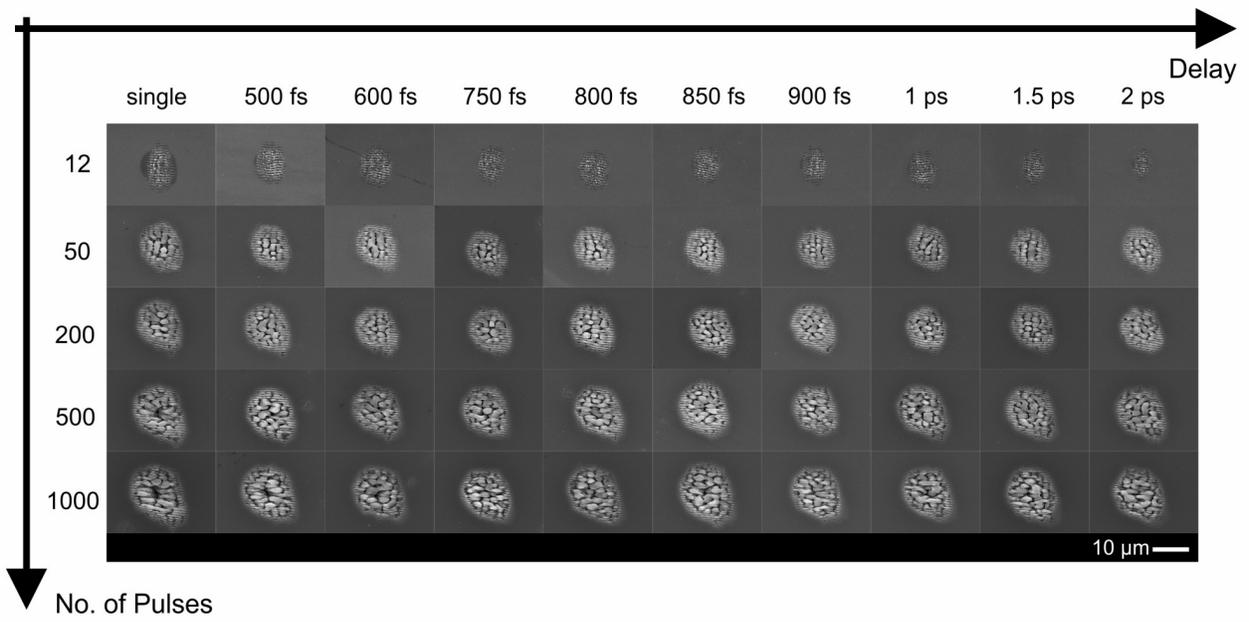

figure2

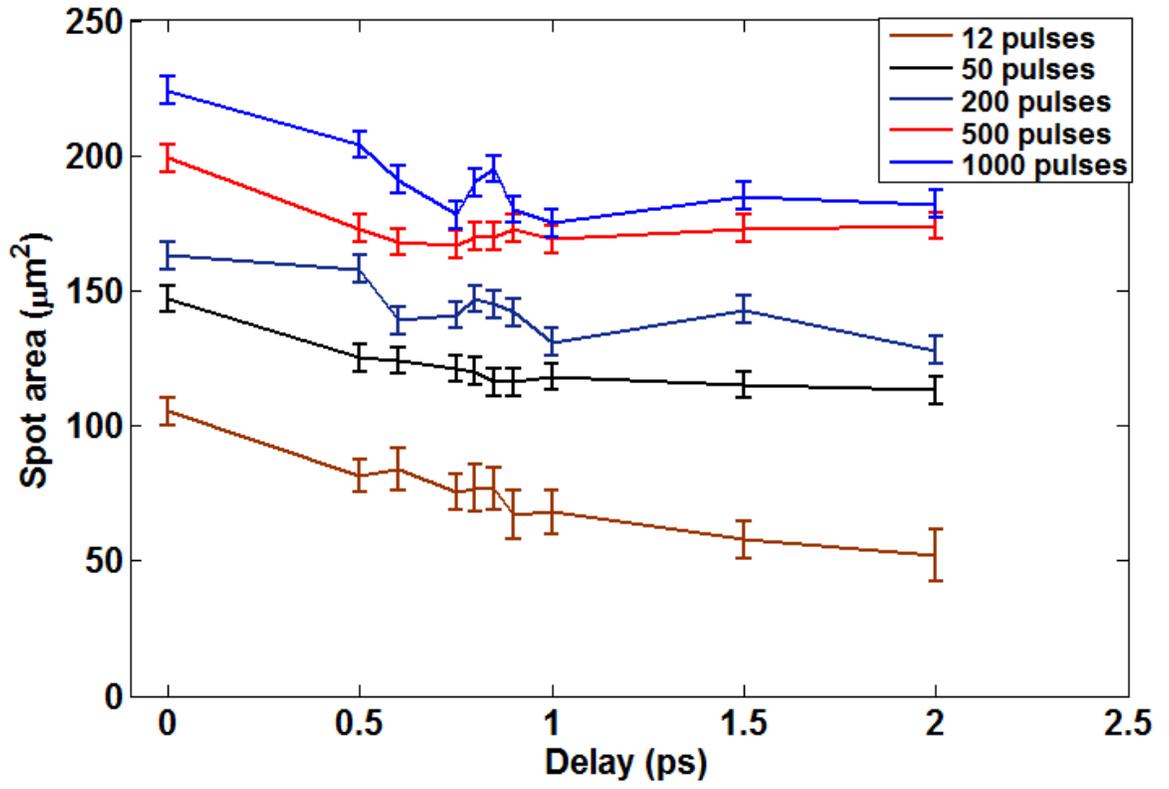

figure3

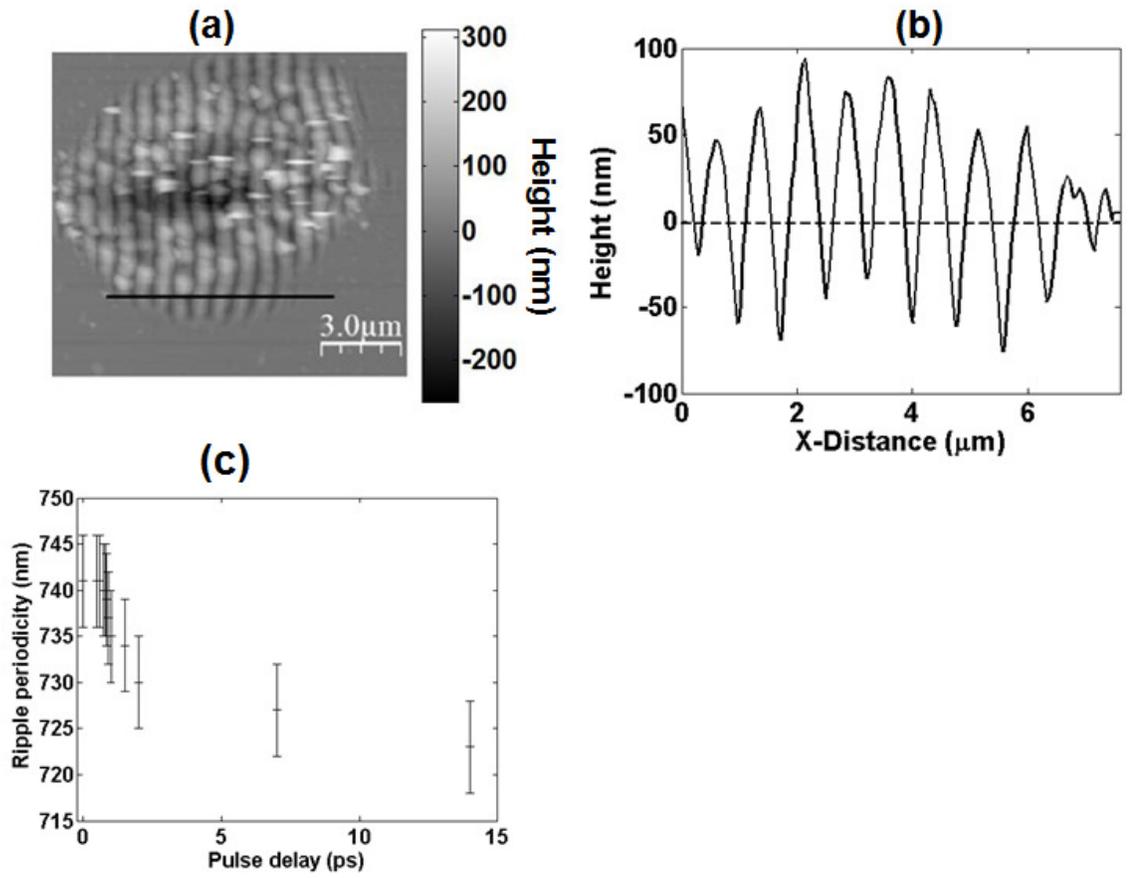

figure4

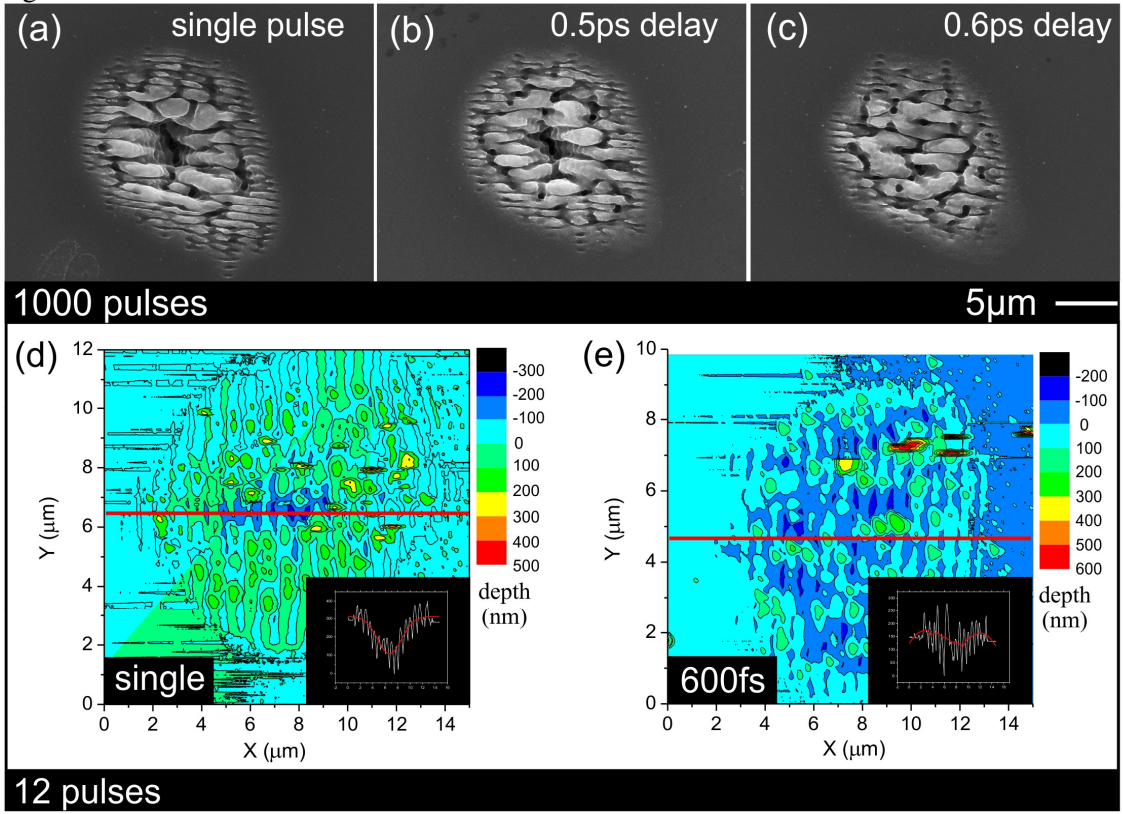

Fig 5

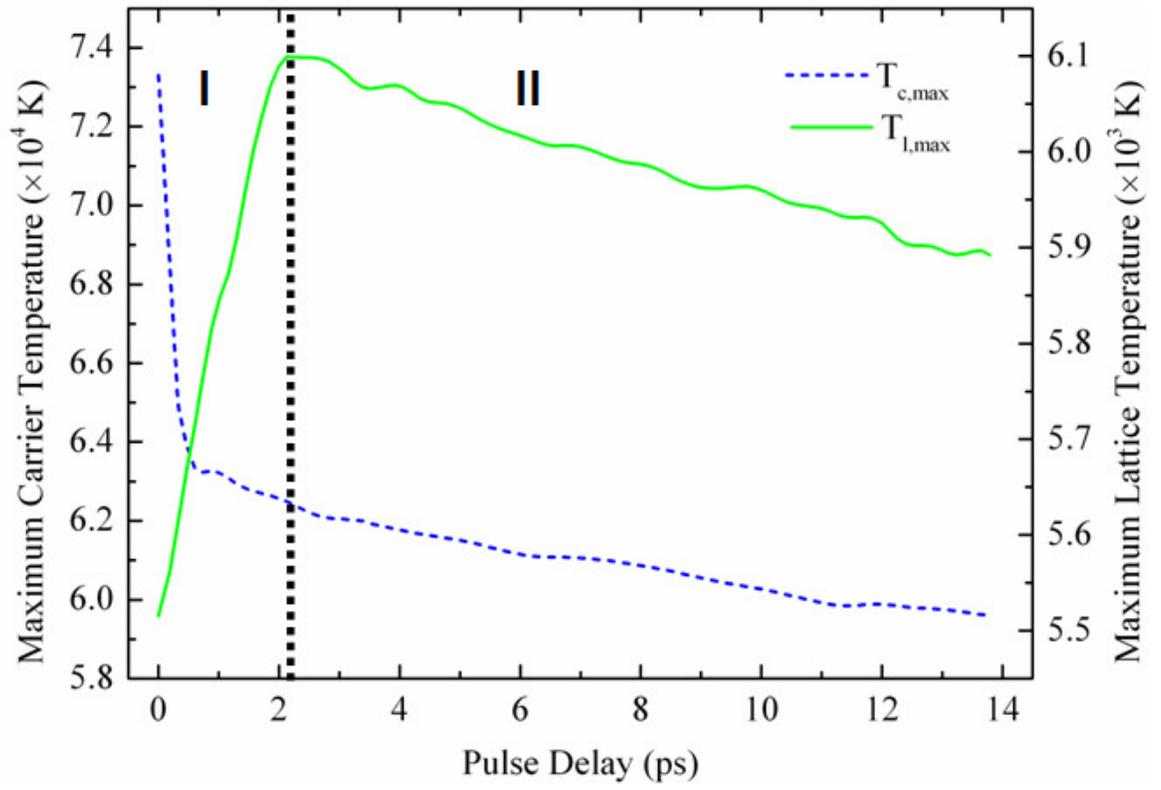

Fig 6

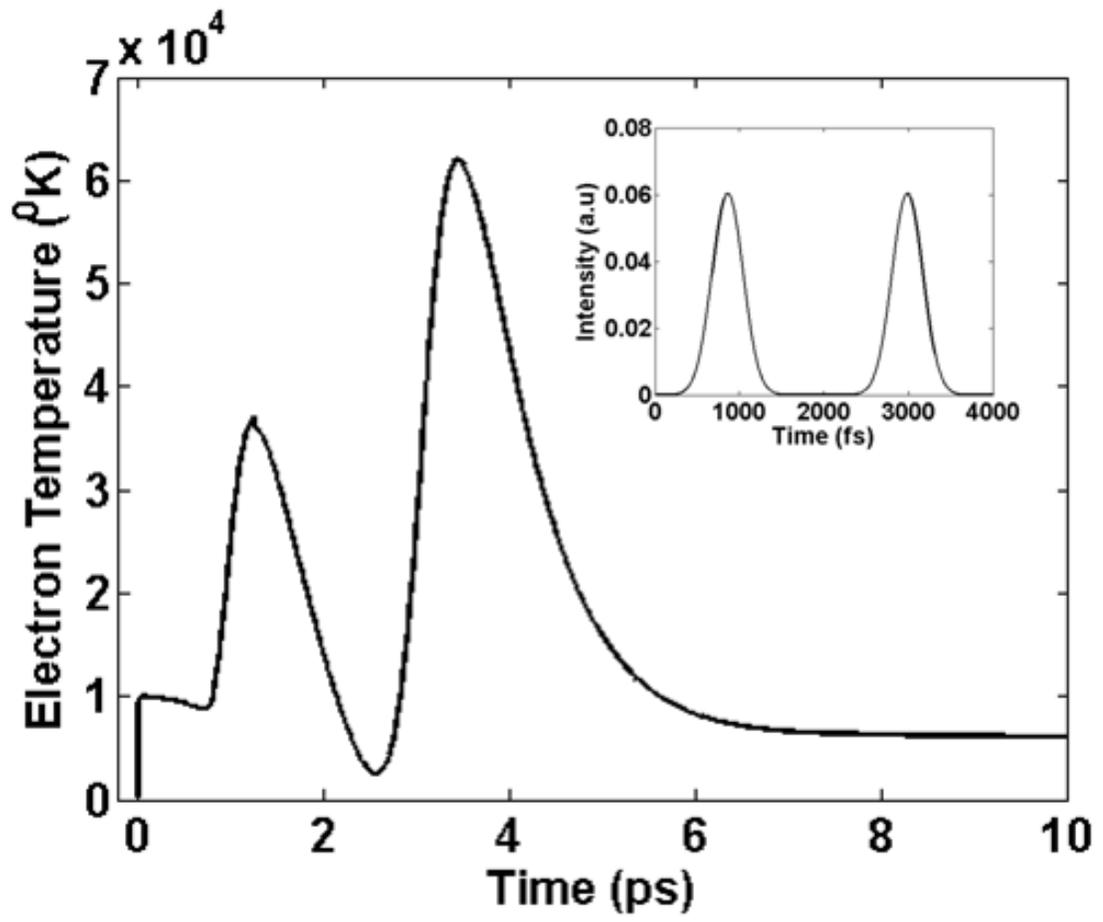

Fig7

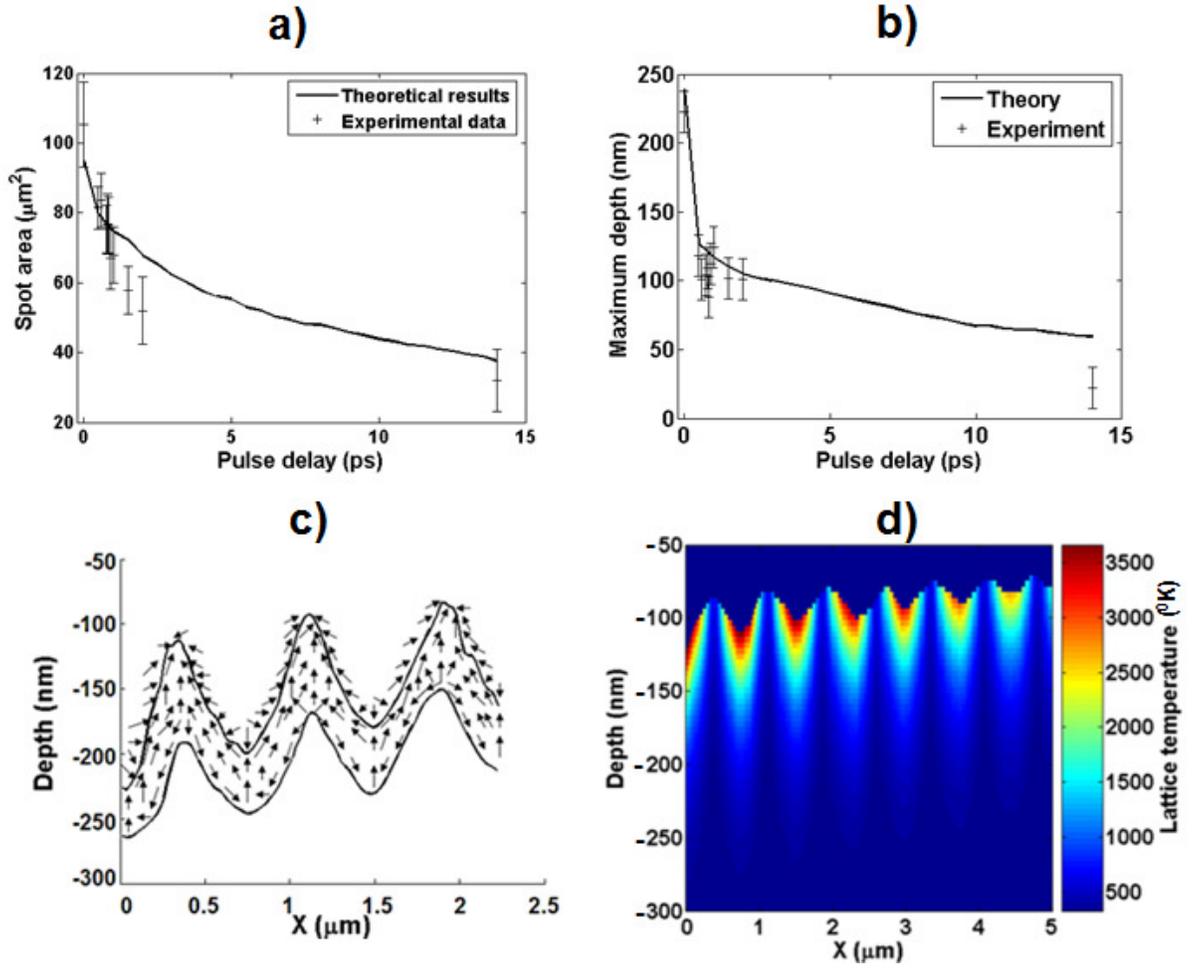

Fig.8

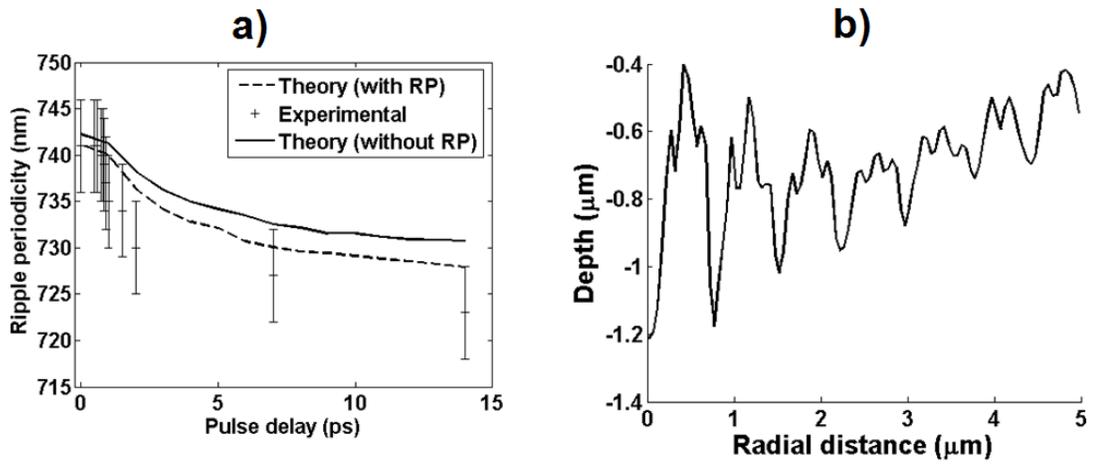